\documentclass[twoside,fleqn]{article}
\usepackage{espcrc2}
\usepackage[dvips]{graphics}

\newcommand{\be}{\begin{equation}}
\newcommand{\ee}{\end{equation}}
\newcommand{\bea}{\begin{eqnarray}}
\newcommand{\eea}{\end{eqnarray}}
\newcommand{\Tr}{\mbox{Tr}}
\newcommand{\bc}{\begin{center}}
\newcommand{\ec}{\end{center}}
\newcommand{\R}{\vec{R}}

\title{String Breaking in Quenched QCD}

\author{Chris Stewart and Roman Koniuk,\\ 
        Physics and Astronomy,
        York University, Toronto, Ontario, M3J 1P3 Canada.}
       
\begin{document}

\begin{abstract}
We present preliminary quenched results on a new operator for the 
investigation of string-breaking within SU(2)-colour QCD. The ground-state 
of a spatially-separated static-light meson-antimeson pair is a combination 
of a state with two distinct mesons, expected to dominate for large 
separations, and a state where the light-quarks have annihilated, which 
contributes for short distances. The crossover between these two regimes 
provides a measure of the string-breaking scale length.
\end{abstract}

\maketitle

\section{INTRODUCTION}

A first-principles demonstration of string-breaking has eluded lattice QCD 
practitioners for many
years. Traditionally, researchers have performed Wilson loop simulations in
full QCD, searching for a plateau in the static quark potential that would
signal the onset of string-breaking. The lack of evidence has led some to
suggest that the Wilson loop operator may have too small an overlap with the
broken two-meson state \cite{de Forcrand,Gusken}, and to recommend a 
search for better operators.

In this paper, we suggest such an operator, consisting of a product of 
spatially-separated static-light meson operators, that provides 
superior overlap with the broken-string state. We also present preliminary 
results that indicate this operator does induce 
string-breaking, even within the quenched approximation. 

\section{THE OPERATOR}

The standard operator used in string-breaking investigations is the 
Wilson loop,
the propagator for a spatially-separated static quark-antiquark pair.
Obviously this operator has strong overlap with the unbroken state of
two static quarks joined by a gluon flux tube---sadly, it has proven to have
insufficient overlap with the broken state of two distinct static-light mesons.

Consider then, a composite operator consisting of a static-light 
meson-antimeson pair, separated by a distance $\R$,
\be
\label{op}
{\cal O}_{M\bar{M}}(\R) = \bar{\psi}_l (0) \gamma_5 \psi_S (0) 
\bar{\psi}_S (\R) \gamma_5 \psi_l (\R) \,.
\ee
The meson-pair propagator is
\be
\label{mmbarp}
G_{M\bar{M}}(t,\R) = {\cal G}_{D} + {\cal G}_{E} \, ,
\ee 
where
\bea
{\cal G}_{D}(t,\R) &=& \Tr \left [ G_{h}(0,t;0,0) \, G^{\dag}_{l}(0,t;0,0) 
\right ] \nonumber \\
& \times &  \Tr \left [ G_{l}(\R,t;\R,0) \, G^{\dag}_{h}(\R,t;\R,0) \right ] 
\nonumber \,, \\
{\cal G}_{E}(t,\R) &=& - \Tr \left [ G_{h}(0,t;0,0) \, 
G^{\dag}_{l}(\R,t;0,0) \right .
\nonumber \\
& & \left .G^{\dag}_{h}(\R,t;\R,0) \, G_{l}(0,t;\R,0) \right ] \, .
\eea
Contributions to ${\cal G}_{D}$ and ${\cal G}_{E}$, the `direct' and 
`exchange' terms, are depicted in Figure 1. 

The large-$R$ limit of this system must be a state with two distinct mesons,
and so we expect ${\cal G}_D$ will dominate for large separations. For small 
$R$, however, the light quarks can easily annihilate, as shown in Figure 1,
leaving a static quark-antiquark pair interacting through the gluon field.
The exchange term ${\cal G}_E$ should contribute strongly for small 
separations, 
giving the necessary overlap with the unbroken state before string-breaking
occurs. 

\begin{figure}[htb]
\begin{center}
\vspace{9pt}
\scalebox{0.6}[0.6]{\includegraphics{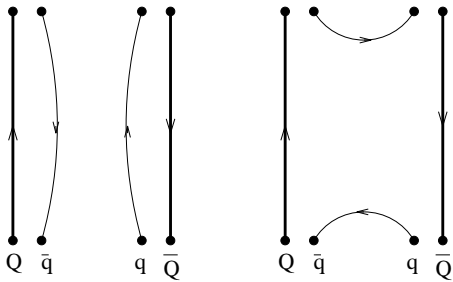}}
\caption{Contributions to direct and exchange terms---note the annihilation
of light quarks in the exchange diagram.}
\end{center}
\end{figure}

\section{THE SIMULATION}

We performed a preliminary simulation to test the validity of the $M\bar{M}$
operator. An ensemble of 150 quenched SU(2)-colour gauge configurations was
created using an ${\cal O}(a^2)$-improved action,
\be
S_{G} = -\beta \sum_{x,\mu > \nu} \left (\frac{5}{3}\frac{P_{\mu\nu}}
{u_0^4} - \frac{1}{12} \frac{R_{\mu\nu} + R_{\nu\mu}}{u_0^6} \right ) \, .
\ee
where $P_{\mu\nu}$ is the plaquette operator, $R_{\mu\nu}$ is a $2\times 1$
loop with the long side along the $\mu$-direction. The lattice dimensions were
$(L_x,L_y,L_z,L_t) = (10,8,8,12)$, with the separation between the static
quarks along the x-direction. 
The tadpole correction $u_0$ was derived from the plaquette operator,
\be
u_0 = \langle P_{\mu\nu} \rangle^{1/4}\, .
\ee
The simulation was performed at
$\beta = 1.07$, corresponding to a lattice spacing of roughly $0.2 fm$, 
using the $\rho$- and $\pi$-meson mass ratio to set the scale.

We used the tadpole-improved Sheikholeslami-Wohlert operator for the
fermion action,
\be
M_{SW} = m_0 + \sum_{\mu} \left (\gamma_{\mu} \triangle_{\mu} - 
\frac{1}{2}
\triangle^2_{\mu}\right ) - \frac{1}{4} \sigma \cdot F \, .
\ee
Due to the computational complexity of this problem, the simulation was 
performed at a very high quark mass, $\kappa = 0.135$, corresponding to a 
pion to rho-meson mass ratio of $m_{\pi}/m_{\rho} \simeq 0.75$. 

\section{RESULTS}

Figure 2 shows a comparison of the direct and exchange
terms in the propagator (\ref{mmbarp}) for varying separation $R$. Also shown
is  a linear-plus-coulomb fit to the Wilson loop data from the same lattice
ensemble. 
Note that for small
R, the exchange term gives a contribution almost identical in size to the 
Wilson loop potential, indicating that the light quarks are indeed annihilating
to leave a static quark-antiquark pair.

\begin{figure}[tb]
\begin{center}
\vspace{9pt}
\scalebox{0.33}[0.33]{\includegraphics{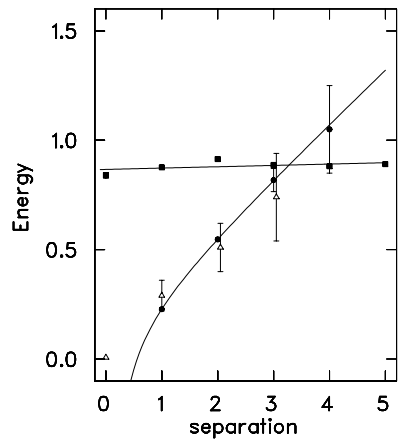}}
\caption{Direct term (squares),
exchange term (triangles) and Wilson loop potential 
(circles and solid line). Exchange term data are 
offset for clarity.}
\end{center}
\end{figure}

We expect string-breaking to occur at the point where the two-meson state 
becomes  energetically
favourable---from the slope of the Wilson loop potential,
this appears to be between $R = 3$ and $R = 4$.
Unfortunately, noise overcomes the exchange term signal just at this point, 
and the crossover can only be inferred from the Wilson loop data. 

The full propagator, shown in Figure 3, gives a
much clearer picture. The system's energy increases to a plateau at the
expected level of the mass of the two separate mesons. The small error bars
for large values of $R$ indicate that, although noise has destroyed the 
exchange term's signal, the mixing into this term has vanished.

\begin{figure}[htb]
\begin{center}
\vspace{9pt}
\scalebox{0.33}[0.33]{\includegraphics{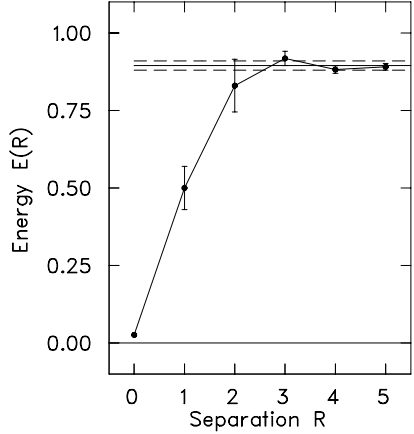}}
\caption{Potential from full propagator. Horizontal line is mass of free
static-light meson.}
\end{center}
\end{figure}

\begin{figure}[htb]
\begin{center}
\vspace{9pt}
\scalebox{0.33}[0.33]{\includegraphics{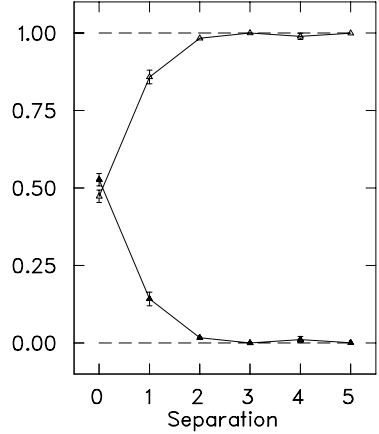}}
\caption{Overlap with direct term (open) and exchange term (closed).}
\end{center}
\end{figure}

Figure 4 confirms this to be the case. The comparative sizes of 
${\cal G}_D(t = 0)$ and
${\cal G}_E(t = 0)$ are taken as a rough measure of the mixing into each of the
direct and exchange terms, and are plotted as a function of $R$. As expected,
the direct and exchange terms both contribute for small $R$, 
and the exchange term vanishes quickly as separation
increases, leaving the direct term to dominate completely for large $R$.

\section{CONCLUSIONS}

The use of new operators, or combinations of operators, to demonstrate
string-breaking on the lattice appears to be an idea whose time has 
arrived---witness the flood of authors in these proceedings presenting results
in $SU(2)$ QCD with scalar fields \cite{Trottier,Philipsen,Knechtli}, 
and references to ongoing research in full $SU(3)$ QCD \cite{Pennanen}. 

We have described an operator suitable for use in string-breaking 
investigations, combining the short-range and long-range physics into a single
operator---that of a spatially separated staic-light meson pair. 
For small separations, this operator behaves like a Wilson loop, resulting in 
a confining potential. For larger separations, the potential reaches a 
plateau at the energy of two distinct static-light mesons, the 
fingerprint of the elusive broken string. 
This occurs even in a {\em quenched} simulation, since the
operator {\em must} energetically favour the meson-pair state for large $R$, 
and so forces the gluon string to break. The light quarks are
providing, within the quenched approximation, some sea-quark effects.

While we are acutely aware of the
limitations of the results described here, we are confident that this operator
will allow accurate determination of the string-breaking distance when used
in more ambitious simulations.

We thank Howard Trottier and Norm Shakespeare for helpful discussions and 
suggestions. This work was supported in part by the National Sciences and 
Engineering Research Council of Canada.

\end{document}